\PassOptionsToPackage{unicode}{hyperref}
\PassOptionsToPackage{hyphens}{url}
\PassOptionsToPackage{dvipsnames,svgnames,x11names}{xcolor}
\documentclass[
]{article}
\usepackage{xcolor}
\usepackage[margin=1in]{geometry}
\usepackage{amsmath,amssymb}
\usepackage{iftex}
\ifPDFTeX
  \usepackage[T1]{fontenc}
  \usepackage[utf8]{inputenc}
  \usepackage{textcomp} 
\else 
  \usepackage{unicode-math} 
  \defaultfontfeatures{Scale=MatchLowercase}
  \defaultfontfeatures[\rmfamily]{Ligatures=TeX,Scale=1}
\fi
\usepackage{lmodern}
\ifPDFTeX\else
\fi
\IfFileExists{upquote.sty}{\usepackage{upquote}}{}
\IfFileExists{microtype.sty}{
  \usepackage[]{microtype}
  \UseMicrotypeSet[protrusion]{basicmath} 
}{}
\makeatletter
\@ifundefined{KOMAClassName}{
  \IfFileExists{parskip.sty}{%
    \usepackage{parskip}
  }{
    \setlength{\parindent}{0pt}
    \setlength{\parskip}{6pt plus 2pt minus 1pt}}
}{
  \KOMAoptions{parskip=half}}
\makeatother
\makeatletter
\ifx\paragraph\undefined\else
  \let\oldparagraph\paragraph
  \renewcommand{\paragraph}{
    \@ifstar
      \xxxParagraphStar
      \xxxParagraphNoStar
  }
  \newcommand{\xxxParagraphStar}[1]{\oldparagraph*{#1}\mbox{}}
  \newcommand{\xxxParagraphNoStar}[1]{\oldparagraph{#1}\mbox{}}
\fi
\ifx\subparagraph\undefined\else
  \let\oldsubparagraph\subparagraph
  \renewcommand{\subparagraph}{
    \@ifstar
      \xxxSubParagraphStar
      \xxxSubParagraphNoStar
  }
  \newcommand{\xxxSubParagraphStar}[1]{\oldsubparagraph*{#1}\mbox{}}
  \newcommand{\xxxSubParagraphNoStar}[1]{\oldsubparagraph{#1}\mbox{}}
\fi
\makeatother

\usepackage{longtable,booktabs,array}
\usepackage{calc} 
\usepackage{etoolbox}
\makeatletter
\patchcmd\longtable{\par}{\if@noskipsec\mbox{}\fi\par}{}{}
\makeatother
\IfFileExists{footnotehyper.sty}{\usepackage{footnotehyper}}{\usepackage{footnote}}
\makesavenoteenv{longtable}
\usepackage{graphicx}
\makeatletter
\newsavebox\pandoc@box
\newcommand*\pandocbounded[1]{
  \sbox\pandoc@box{#1}%
  \Gscale@div\@tempa{\textheight}{\dimexpr\ht\pandoc@box+\dp\pandoc@box\relax}%
  \Gscale@div\@tempb{\linewidth}{\wd\pandoc@box}%
  \ifdim\@tempb\p@<\@tempa\p@\let\@tempa\@tempb\fi
  \ifdim\@tempa\p@<\p@\scalebox{\@tempa}{\usebox\pandoc@box}%
  \else\usebox{\pandoc@box}%
  \fi%
}
\def\fps@figure{htbp}
\makeatother

\NewDocumentCommand\citeproctext{}{}
\NewDocumentCommand\citeproc{mm}{%
  \begingroup\def\citeproctext{#2}\cite{#1}\endgroup}
\makeatletter
 \let\@cite@ofmt\@firstofone
 \def\@biblabel#1{}
 \def\@cite#1#2{{#1\if@tempswa , #2\fi}}
\makeatother
\newlength{\cslhangindent}
\newlength{\csllabelwidth}
\newenvironment{CSLReferences}[2] 
 {\begin{list}{}{%
  \setlength{\itemindent}{0pt}
  \setlength{\leftmargin}{0pt}
  \setlength{\parsep}{0pt}
  \ifodd #1
   \setlength{\leftmargin}{\cslhangindent}
   \setlength{\itemindent}{-1\cslhangindent}
  \fi
  \setlength{\itemsep}{#2\baselineskip}}}
 {\end{list}}
\usepackage{calc}

\usepackage{pdflscape}
\usepackage{booktabs}
\usepackage{longtable}
\usepackage{array}
\usepackage{multirow}
\usepackage{float}
\makeatletter
\@ifpackageloaded{caption}{}{\usepackage{caption}}
\AtBeginDocument{%
\ifdefined\contentsname
  \renewcommand*\contentsname{Table of contents}
\else
  \newcommand\contentsname{Table of contents}
\fi
\ifdefined\listfigurename
  \renewcommand*\listfigurename{List of Figures}
\else
  \newcommand\listfigurename{List of Figures}
\fi
\ifdefined\listtablename
  \renewcommand*\listtablename{List of Tables}
\else
  \newcommand\listtablename{List of Tables}
\fi
\ifdefined\figurename
  \renewcommand*\figurename{Figure}
\else
  \newcommand\figurename{Figure}
\fi
\ifdefined\tablename
  \renewcommand*\tablename{Table}
\else
  \newcommand\tablename{Table}
\fi
}
\@ifpackageloaded{float}{}{\usepackage{float}}
\floatstyle{ruled}
\@ifundefined{c@chapter}{\newfloat{codelisting}{h}{lop}}{\newfloat{codelisting}{h}{lop}[chapter]}
\floatname{codelisting}{Listing}

\makeatother
\makeatletter
\@ifpackageloaded{caption}{}{\usepackage{caption}}
\@ifpackageloaded{subcaption}{}{\usepackage{subcaption}}
\makeatother
\usepackage{bookmark}
\IfFileExists{xurl.sty}{\usepackage{xurl}}{} 
\hypersetup{
  pdftitle={Open-Weight LLMs Are Often Competitive with Commercial APIs for Political Science Text Classification},
  pdfauthor={Hanno Hilbig},
  colorlinks=true,
  linkcolor={blue},
  filecolor={Maroon},
  citecolor={Blue},
  urlcolor={Blue},
  pdfcreator={LaTeX via pandoc}}

\title{Open-Weight LLMs Are Often Competitive with Commercial APIs for
Political Science Text Classification}
\author{Hanno
Hilbig\thanks{hhilbig@ucdavis.edu. Code and data: \url{https://github.com/hhilbig/polsci-open-bench}.}}
\date{2026-05-18}
\begin{document}
\maketitle

\subsection*{Abstract}\label{abstract}
\addcontentsline{toc}{subsection}{Abstract}

Can researchers use local open-weight models instead of commercial APIs
for LLM text classification? Local models avoid marginal API charges,
keep data on the researcher's machine, and make exact model versions
easier to preserve. I benchmark five local models against four
commercial API models on 34 political science classification tasks.
Local models are often competitive, especially on simpler tasks. In a
task-specific oracle comparison, local models match or exceed API
performance on 9 tasks; on average, the best API model exceeds the best
local model by 0.015 F1. The four strongest observed model means fall
within 0.021 F1. API models have their clearest edge on complex tasks
with many labels or multiple outputs per item. Batching several items in
one prompt usually reduces local runtime per item, but specific
model-task pairs can return invalid response formats or labels. Taken
together, the results make local open-weight models a practical
candidate alternative for many political science classification tasks,
provided researchers validate candidate models on task-specific labels
and check batching reliability before scaling up.

\section{Motivation}\label{motivation}

When using LLMs to classify text data, researchers can use commercial
API models or local open-weight models. Commercial API models are
typically reliable and easy to call from R or Python. However, they cost
money per API call, send data off-machine, and create reproducibility
risk when model behavior changes over time. The alternative is local
open-weight models. These models avoid marginal API charges, keep the
data on the researcher's machine, and make exact model versions easier
to preserve and track. This matters when researchers work with
restricted data or when data-use rules prohibit third-party APIs.
Open-weight models can also be useful when research funds are not
abundant, or for exploratory tasks that may otherwise not be pursued
because of costs.

This benchmark focuses on three questions: whether local models approach
commercial API performance, how long local classification takes, and
when task complexity widens the API advantage. Drawing on 34 tasks from
political science papers, public replication archives, and documented
public datasets, I cover the kinds of coding problems applied
researchers routinely face: relevance, stance, tone, events, claims,
relations, issues, and topics.

The main result is practical rather than a single leaderboard. Local
models are often close to API models, especially on simpler tasks, but
performance varies enough across tasks that researchers should validate
candidate models on the target coding problem. APIs retain their
clearest edge on more complex tasks. Prompt batching can reduce local
runtime, but only after checking invalid-output rates.

\section{Design}\label{design}

The benchmark has two parts. First, I run a one-item-at-a-time benchmark
covering thirty-four classification tasks drawn from political science
papers, public replication archives, and documented public datasets. I
compare nine models: five local open-weight models and four commercial
API models from OpenAI, Anthropic, and DeepSeek. This produces 306
model-task comparisons. Each model sees the same items within a task.
Thirty-one tasks use 500 sampled items. Three tasks have fewer cleaned
items available, so I use all available items: 320 for Brandt political
relevance, 312 for PLOVER event coding, and 293 for COVID threat
minimization. This is why task sizes range from 293 to 500 items.

I report performance and reliability for each task and model. The main
F1 score puts precision and recall on the same scale. For binary tasks,
I use the F1 score for the positive class, the harmonic mean of
precision and recall. For multi-binary tasks, I average the per-label F1
scores. For single-label categorical tasks, I use macro F1, which gives
each observed class equal weight before averaging. Where a task has one
output per item, I also report accuracy and Matthews correlation
coefficient (MCC). The multi-binary ATI task is excluded from mean
accuracy and MCC because exact-match accuracy and multilabel MCC would
answer a different question. This metric choice matters because class
imbalance is common: macro-style F1 highlights rare-class performance,
accuracy gives the plain success rate, and MCC is less sensitive than
accuracy to majority-class dominance.

I also record time per item and the share of unusable responses. A
response is unusable if it fails to parse, returns the wrong object
shape, or uses a label outside the allowed label set. I compute
performance metrics over usable outputs only, so failure rates need to
be read alongside F1, accuracy, and MCC. In the one-item-at-a-time
benchmark, 95 of 147,825 outputs are unusable (0.064\%). The issue is
therefore small for the main performance comparisons, but not zero.

I evaluate each model-item pair once in the one-item-at-a-time
benchmark. Table \ref{tab:models} reports the exact model IDs, local
model tags, and runtime details. The local setup was a single Apple
Silicon MacBook with 32 GB of unified memory running Ollama at
temperature 0.1, \texttt{num\_predict=1024}, and thinking disabled.
Local calls used prompt-only JSON instructions, after which the runner
parsed and validated labels against the task schema. Commercial API
calls used provider-specific constrained-output settings. OpenAI tiers
used \texttt{reasoning\_effort=medium}, strict JSON schema, and
\texttt{max\_completion\_tokens=2000}; DeepSeek used JSON object mode,
disabled thinking, \texttt{max\_tokens=1024}, and temperature 0.1; and
Anthropic used forced tool use with \texttt{max\_tokens=2000}. Other
than provider SDK retries for transient request failures, I do not retry
unusable outputs in the one-item-at-a-time benchmark. I use the source
paper's coding instructions verbatim where possible; otherwise, I derive
prompts from source codebooks, public label definitions, or task
descriptions (see Table \ref{tab:tasks}). Source labels in the task
table correspond to full bibliography entries.

\begin{table}[H]
\centering
\caption{Models and runtime settings. The table reports exact local Ollama tags and API model IDs used in the benchmark. No context-window override was set; output caps and output constraints are described in the text. \label{tab:models}}
\centering
\resizebox{\ifdim\width>\linewidth\linewidth\else\width\fi}{!}{
\fontsize{8}{10}\selectfont
\begin{tabular}[t]{ll>{\raggedright\arraybackslash}p{15em}l>{\raggedright\arraybackslash}p{16em}}
\toprule
Model & Class & Exact ID/tag & Size/quant. & Access/runtime\\
\midrule
Gemma 4 26B & Local Ollama & gemma4:26b & 26B, Ollama tag default & Apple M2 Pro, 32 GB; Ollama 0.19.0; macOS 26.1\\
Gemma 4 31B & Local Ollama & gemma4:31b-it-q4\_K\_M & 31B, Q4\_K\_M & Apple M2 Pro, 32 GB; Ollama 0.19.0; macOS 26.1\\
Qwen3 14B & Local Ollama & qwen3:14b-q4\_K\_M & 14B, Q4\_K\_M & Apple M2 Pro, 32 GB; Ollama 0.19.0; macOS 26.1\\
Qwen3 30B-A3B & Local Ollama & qwen3:30b-a3b-q4\_K\_M & 30B MoE, Q4\_K\_M & Apple M2 Pro, 32 GB; Ollama 0.19.0; macOS 26.1\\
Mistral Sm 24B & Local Ollama & mistral-small:24b-instruct-2501-q4\_K\_M & 24B, Q4\_K\_M & Apple M2 Pro, 32 GB; Ollama 0.19.0; macOS 26.1\\
\addlinespace
gpt-5.5 & API & gpt-5.5 & provider model & OpenAI, May 2026\\
gpt-5.4-nano & API & gpt-5.4-nano & provider model & OpenAI, May 2026\\
DeepSeek V4 Pro & API & deepseek-v4-pro & provider model & DeepSeek, May 2026\\
Claude Sonnet 4.6 & API & claude-sonnet-4-6 & provider model & Anthropic, May 2026\\
\bottomrule
\end{tabular}}
\end{table}

Second, I test whether prompt batching speeds up local inference. In the
batching experiment, the model receives ten items in one prompt and
returns ten labels in a single response. This tests whether batching
reduces time per item without changing the classification task. The
local grid with 10 items per prompt covers all thirty-four tasks and all
five local models.

The one-item-at-a-time benchmark is the basis for all main performance
comparisons. I treat prompt batching as a separate speed and reliability
check because it changes the format of the model call and can change
invalid-output rates. Separating the two designs keeps the performance
comparison clean while still showing whether local use is practical at
larger scale.

Finally, I use uncertainty checks to avoid overinterpreting small
differences. Within-task comparisons use 95\% paired-by-item bootstrap
confidence intervals (1000 iterations). I treat the thirty-four tasks as
a fixed benchmark population, but also report paired task-level
bootstrap intervals that resample tasks with replacement.

\section{Results}\label{results}

\subsection{Average performance}\label{average-performance}

\begin{figure}[!h]

\centering{

\includegraphics[width=0.82\linewidth,height=\textheight,keepaspectratio]{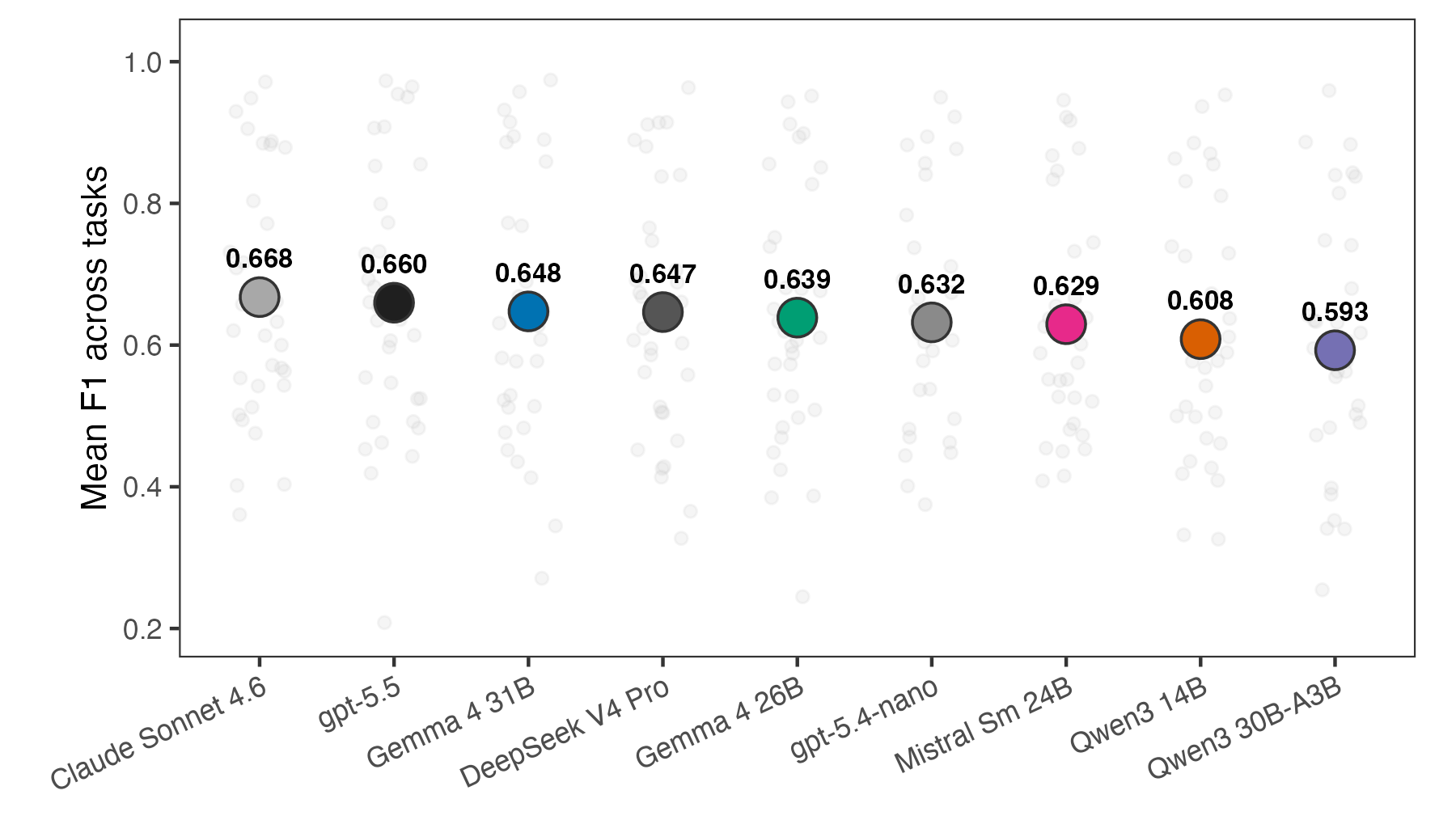}

}

\caption{\label{fig-mean-f1}Mean F1 across thirty-four tasks per model.
Large filled circles show 34-task means. Small gray dots show observed
per-task main F1 values.}

\end{figure}

Figure \ref{fig-mean-f1} shows the 34-task model averages and the spread
of task-level performance within each model. Claude Sonnet 4.6 has the
highest observed mean F1 (0.668), followed by gpt-5.5 (0.660), Gemma 4
31B (0.648), and DeepSeek V4 Pro (0.647). The four strongest observed
model means fall within 0.021 F1. The same broad top group appears under
mean accuracy and mean MCC where those metrics are defined (Table
\ref{tab:altmetrics}), although the exact order changes slightly. Among
the OpenAI tiers, gpt-5.4-nano remains the cost-efficient option, at
0.632 mean F1 compared with 0.660 for gpt-5.5.

The task-level bootstrap gives the same result (Table
\ref{tab:task-bootstrap}). The Claude Sonnet 4.6 versus gpt-5.5 interval
includes zero, Gemma 4 31B and DeepSeek V4 Pro are effectively tied, and
the estimated Claude advantage over Gemma 4 31B and DeepSeek V4 Pro is
about two F1 points. Many within-task leader differences are also too
uncertain to treat as firm rankings. In practice, these small average
gaps are not enough to rank the models reliably.

\begin{figure}[!htbp]

\centering{

\includegraphics[width=0.74\linewidth,height=\textheight,keepaspectratio]{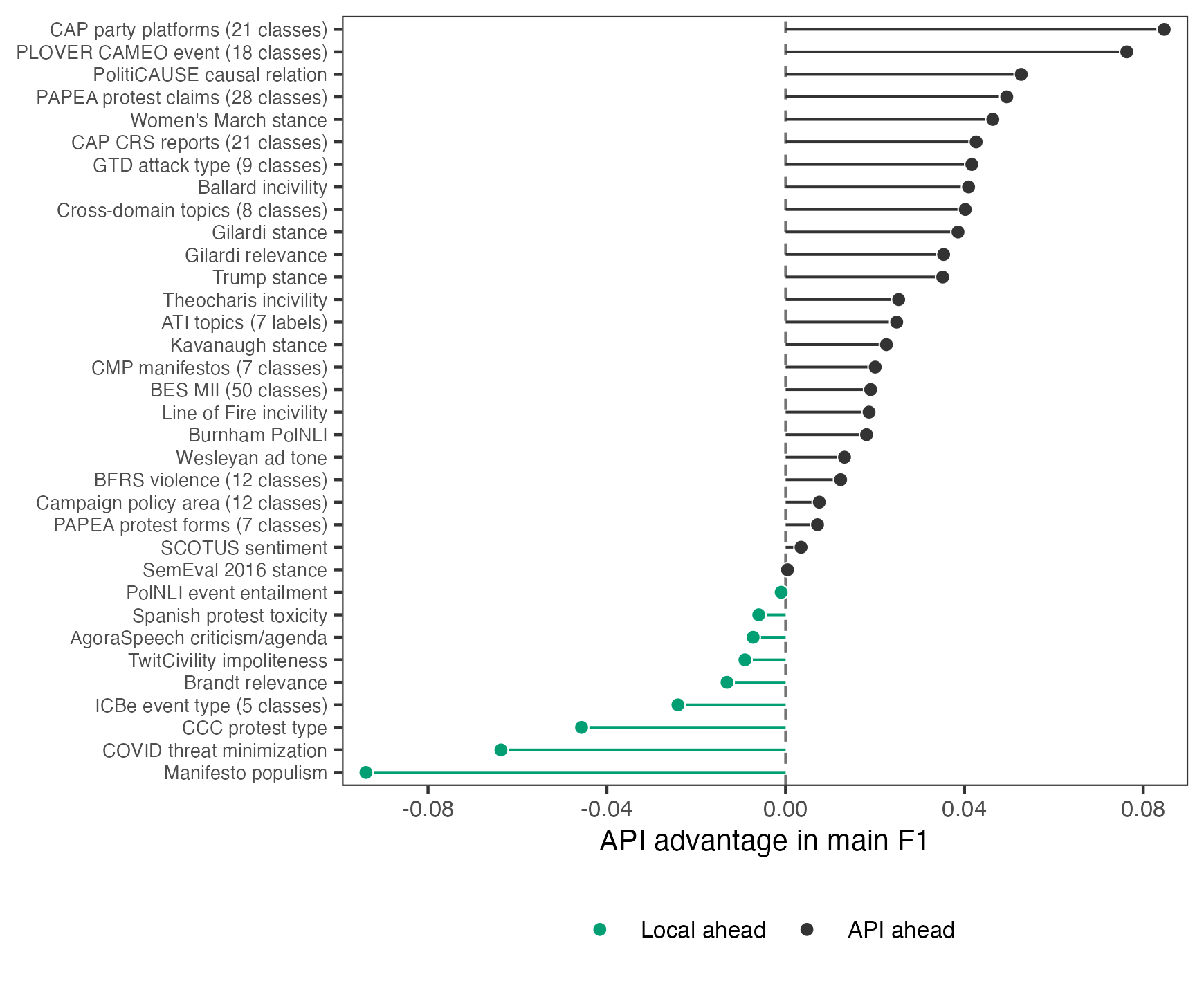}

}

\caption{\label{fig-best-local-api-gap}API advantage in main F1 by task.
Values below zero favor the local model class; values above zero favor
the API model class.}

\end{figure}

The local/API comparison in Figure \ref{fig-best-local-api-gap} directly
addresses the paper's main question. After selecting the best model
within each class for each task, the best local model matches or exceeds
the best API model on 9 of the 34 tasks. On average, the best API model
exceeds the best local model by 0.015 F1, and the median API advantage
is 0.019. Figure \ref{fig-best-local-api-gap} uses the same sign
convention: values below zero favor local models and values above zero
favor API models. This is not evidence that local models dominate API
models. It shows that the API advantage is often small enough that local
model calls are worth validating on the target task.

Table \ref{tab:model-selection-value} reports how much performance would
improve if researchers could choose the best model separately for each
task. These upper bounds choose the best model after observing
performance on each task. Allowing the local model to vary by task
raises mean F1 from 0.648 for the best single local model to 0.673.
Across all nine models, task-specific selection reaches 0.696, compared
with 0.668 for the best single model. These values show why
task-specific validation can be valuable, but they do not show how large
a validation sample is needed to choose models reliably.

The larger differences are across tasks. The paired-by-item intervals
are wide enough that many task-level leader differences remain
uncertain. I therefore do not read the benchmark primarily as a
leaderboard: global rank is a weak decision rule for applied coding.

\subsection{Different models lead on different
tasks}\label{different-models-lead-on-different-tasks}

\begin{figure}[!h]

\centering{

\includegraphics[width=0.86\linewidth,height=\textheight,keepaspectratio]{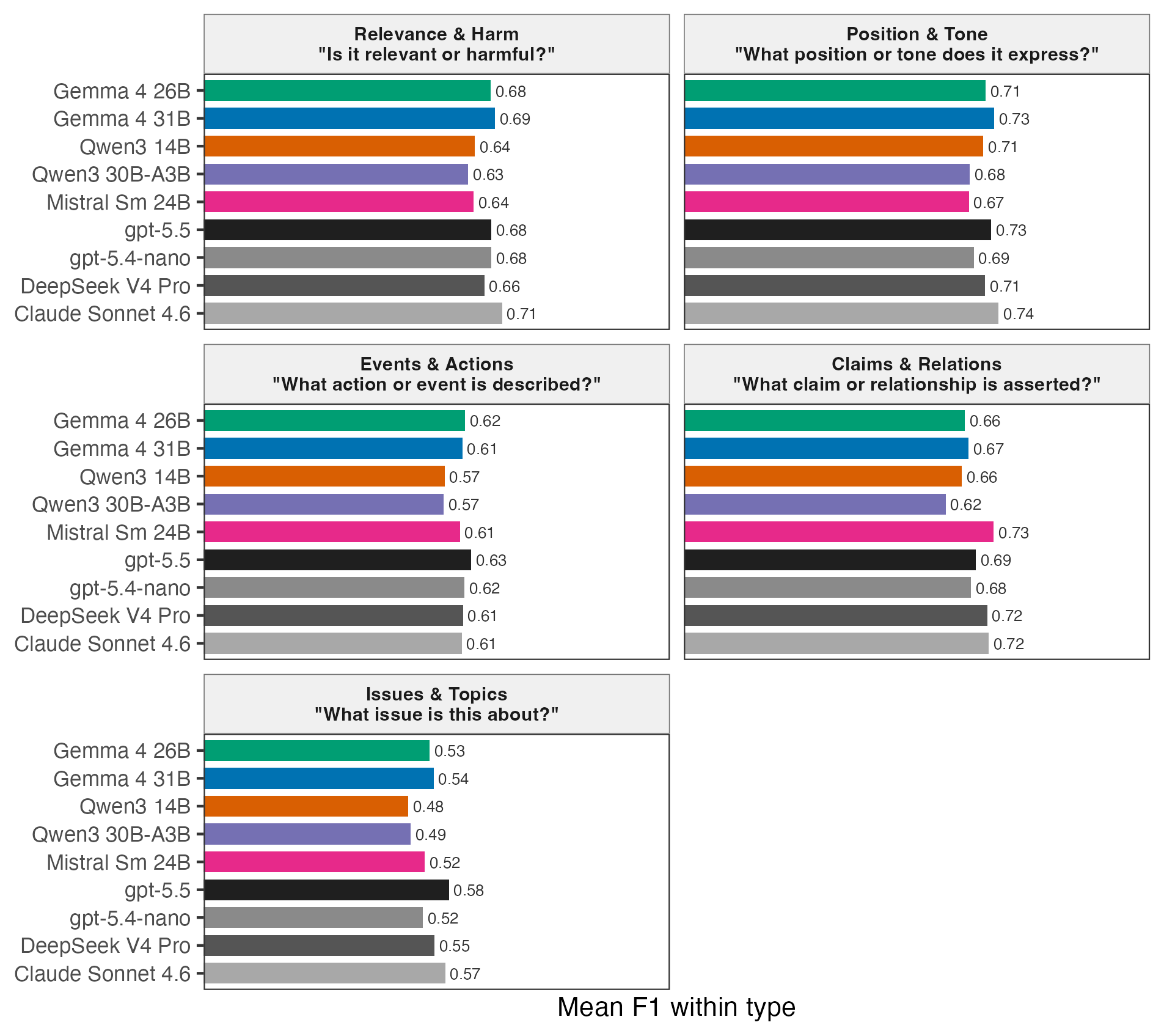}

}

\caption{\label{fig-family}Mean F1 within five broad annotation types,
by model. Type-level averages are descriptive only.}

\end{figure}

Figure \ref{fig-family} groups the thirty-four tasks into five broad
annotation types: relevance and harm, position and tone, events and
actions, claims and relations, and issues and topics. These are my
groupings, not source-provided task families, and I assign borderline
cases by the prediction target. The full task list is in Table
\ref{tab:tasks}. The type-level averages do not show a simple local
versus API pattern.

At the task level, the top point estimates are split across eight models
(Table \ref{tab:winners}). gpt-5.5 leads on thirteen tasks, Claude
Sonnet 4.6 leads on eight, Mistral Small leads on three, and five other
models lead on two tasks each.

The winner split matters because model strengths differ by task. Even
models with lower average performance sometimes produce the best
task-level point estimate. A model that is weaker on average can still
be the best model for a specific task.

\subsection{Performance is lower on more complex coding
tasks}\label{performance-is-lower-on-more-complex-coding-tasks}

\begin{figure}[!h]

\centering{

\includegraphics[width=0.84\linewidth,height=\textheight,keepaspectratio]{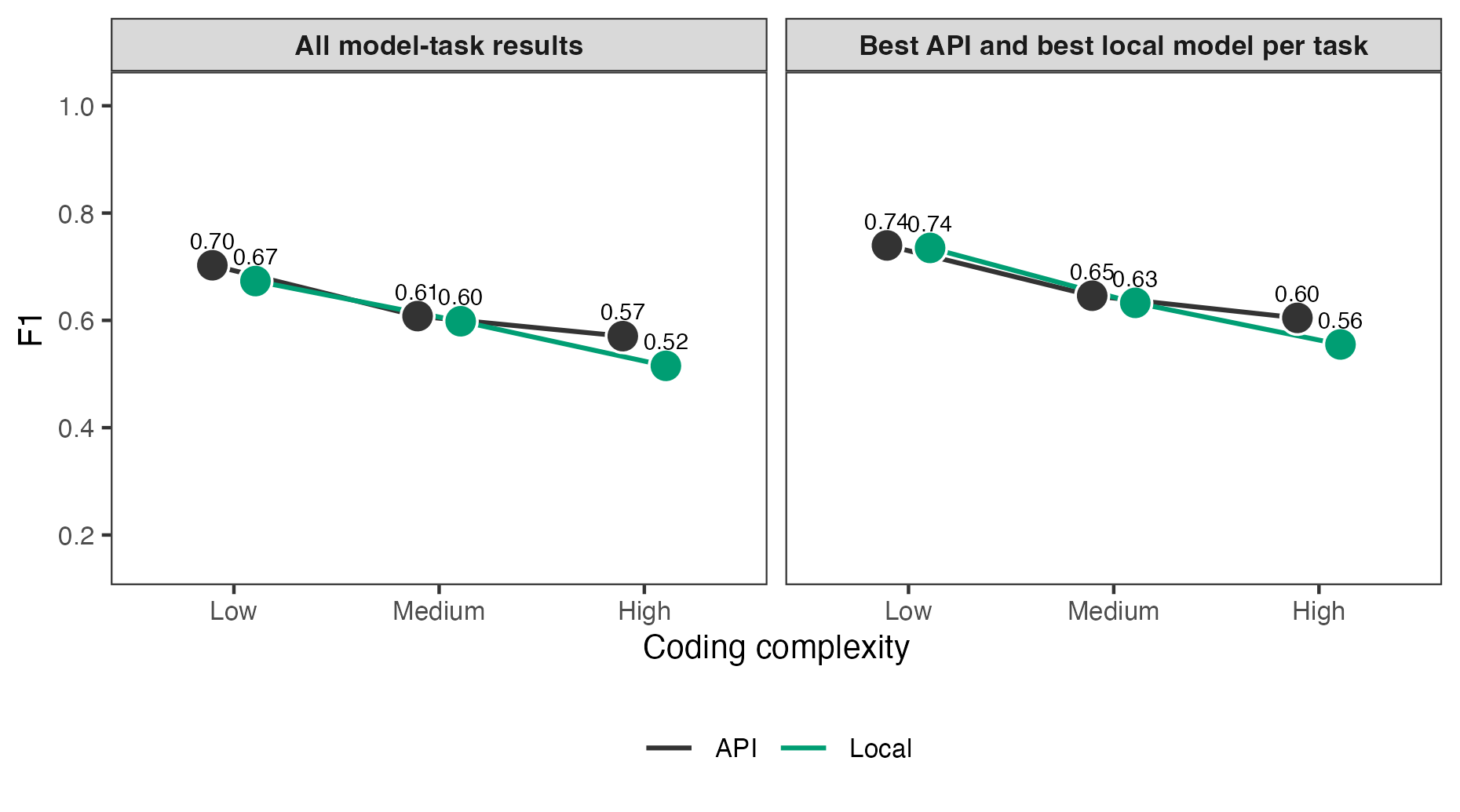}

}

\caption{\label{fig-complexity}Coding complexity and model performance
across the 34-task benchmark. The left panel shows mean F1 across all
model-task results, separately for local and API models. The right panel
shows the best local model and best API model per task within each
complexity group. Points show group means; lines connect complexity
groups within model class.}

\end{figure}

Figure \ref{fig-complexity} examines whether coding complexity is
associated with the large task-to-task variation. I measure complexity
with two features that researchers can observe before running a
benchmark: the prompt/codebook and the number of labels used in the gold
data. Effective labels measure how many labels are actually used in the
gold data, rather than how many labels appear in the codebook.
Low-complexity tasks have fewer than three effective labels and short
prompts. Medium-complexity tasks have either more labels or longer
prompts. High-complexity tasks have at least eight effective labels or
require multi-binary outputs.

Performance falls as coding tasks become more complex, but local models
remain close on many tasks. Mean F1 falls from 0.70 to 0.57 for API
model-task pairs and from 0.67 to 0.52 for local pairs as tasks move
from low to high complexity. When I instead compare the best API model
with the best local model on each task, the low-complexity gap nearly
disappears (0.740 versus 0.735), while the high-complexity gap is about
0.05 F1 (0.605 versus 0.555). This is the clearest API edge in the
complexity groups: long codebooks, many effective labels, and tasks with
multiple outputs per item.

\begin{figure}[!h]

\centering{

\includegraphics[width=0.7\linewidth,height=\textheight,keepaspectratio]{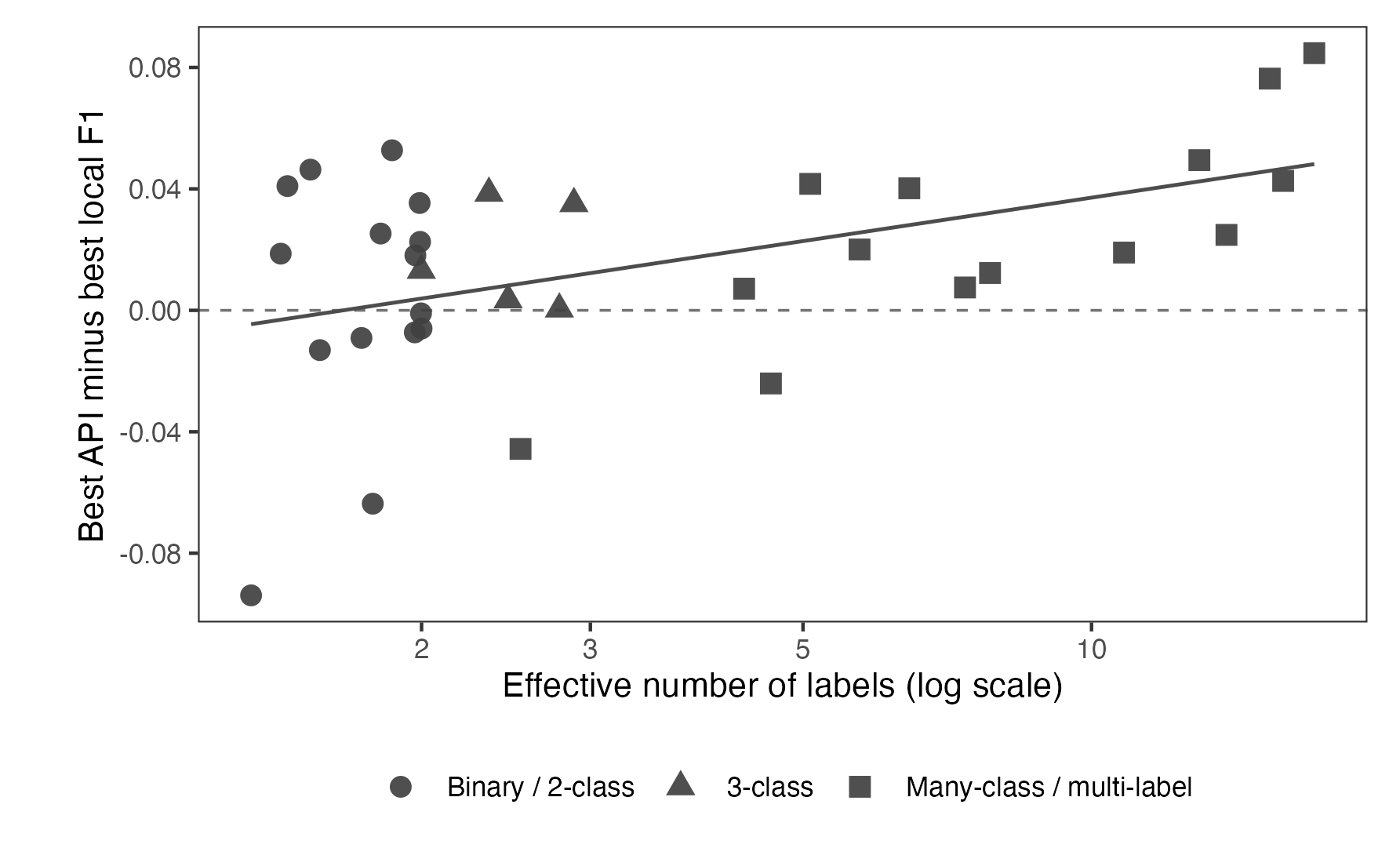}

}

\caption{\label{fig-label-structure-gap}API advantage in main F1 by
effective number of labels. The dashed horizontal line marks equal
performance; the solid line is a descriptive linear fit across tasks.}

\end{figure}

Figure \ref{fig-label-structure-gap} shows a related pattern: tasks with
more actively used labels tend to show larger API advantages. Effective
label count measures how many labels actually matter in the sample,
rather than simply counting every possible label. The relationship is
positive. The simple correlation between effective label count and the
API-minus-local gap is about 0.50. Binary and two-class tasks have a
mean API advantage of 0.004, while many-class or multi-label tasks have
a mean API advantage of 0.025.

\subsection{Local inference is not uniformly
slower}\label{local-inference-is-not-uniformly-slower}

\begin{figure}[!htbp]

\centering{

\includegraphics[width=0.72\linewidth,height=\textheight,keepaspectratio]{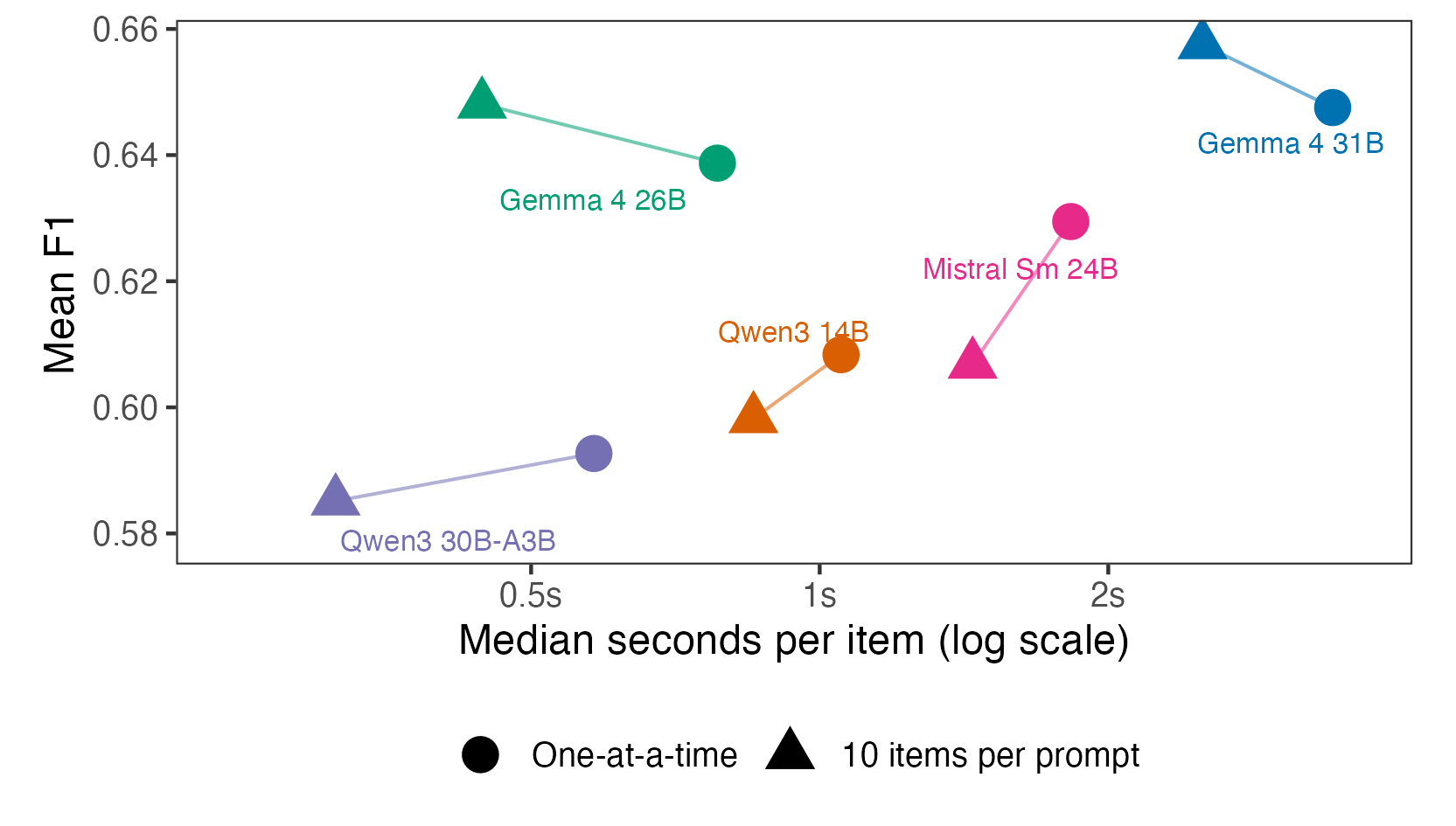}

}

\caption{\label{fig-speed}Mean F1 vs median runtime per item, for the
five local models. Circles show one-at-a-time calls over the 34-task
benchmark; triangles show calls with 10 items per prompt over the same
task set. Lines connect the two call modes for the same model, and
labels identify model pairs. API models are excluded because their
per-call time mixes server load, queueing, and network delay and is not
directly comparable to local inference time.}

\end{figure}

Local inference speed varies sharply with model size, model design, and
batching (Figure \ref{fig-speed}). On a typical 500-item task, median
one-item-at-a-time runtimes range from about 5 minutes for Qwen3-30B-A3B
and 7 minutes for Gemma 4 26B to about 29 minutes for Gemma 4 31B.
Qwen3-30B-A3B is fastest because it is a sparse model that uses about
three billion active parameters per call. Gemma 4 31B has the strongest
local 34-task one-item-at-a-time F1, while Gemma 4 26B is much faster
and remains close on average. Local inference is slow enough to plan
around, but not slow enough to rule out ordinary validation and
production coding.

All five local models ran on a single 32 GB Apple Silicon MacBook; the
largest, Gemma 4 31B, fit in unified memory. Appendix Figure
\ref{fig-local-runtime-per-1000} translates the local runtimes into
median minutes per 1,000 items for each model. For reference, the API
tiers in this run took roughly 0.9 to 1.7 seconds per item end-to-end,
but those numbers mix model speed with provider load, network delays,
and queueing and are not shown in the figure.

\subsection{Batching speeds up local inference but can produce invalid
output}\label{batching-speeds-up-local-inference-but-can-produce-invalid-output}

Prompt batching changes the prompt: the model receives several items at
once and must return one label for each item. Provider Batch API mode is
different. It changes how requests are queued and billed, but each item
still appears in a separate request. I test prompt batching here. Pipal
et al. (\citeproc{ref-pipal2026}{2026}) show why this matters for
labeling tasks: one-at-a-time classification repeats the same prompt and
request setup for every item.

Figure \ref{fig-speed} shows the main local result. Prompt batching
reduces median runtime per item for every local model, with model-level
median speedups from about 1.23× to 1.86×. Among usable outputs, average
F1 changes range from -0.023 to +0.010. When I instead count unusable
outputs as failures, model-level mean F1 changes range from -0.025 to
+0.010 relative to one-item-at-a-time calls. Gemma 4 26B is the most
useful fast batched option in these results: it remains close to Gemma 4
31B on average but classifies a typical 500-item task in about 4 minutes
with 10 items per prompt, compared with about 21 minutes for Gemma 4
31B.

Batching helps less when prompts are long. Short-prompt model-task pairs
have a median local speedup of about 1.34× with 10 items per prompt,
while long-codebook pairs have a median speedup of about 1.20×. The
bigger problem is invalid output: some models return invalid JSON, the
wrong object shape, or labels outside the allowed set. The largest
failures are concentrated in specific model-task pairs. For example,
Gemma 4 26B produces unusable responses for 72\% of Erlich ATI items
with 10 items per prompt, and Qwen3-30B-A3B produces unusable responses
for 51\% of Halterman CCC items with 10 items per prompt. In a separate
10-task prompt-batching diagnostic for OpenAI models, gpt-5.5 returned
unusable output for 68\% of Halterman CCC items. That OpenAI diagnostic
does not enter the main one-item-at-a-time performance results. Table
\ref{tab:malformed} lists the covered response-format and label failure
rates above 5\%.

\subsection{Cost}\label{cost}

The OpenAI portion of the 34-task one-item-at-a-time benchmark remains
cheap enough for benchmarking on a validation sample, but the tiers
differ sharply. At publicly listed prices, gpt-5.4-nano costs about \$4
for the 16,425 predictions in the 34-task benchmark, while gpt-5.5 would
cost about \$71 under direct per-request pricing. In return, gpt-5.5
improves mean F1 from 0.632 to 0.660. The smaller tier is therefore
worth testing before moving to a flagship API model.

Provider Batch API mode changes the cost calculation for flagship
models. This is provider-side request processing, not prompt batching
with several items in one prompt. For a 16-task subset, I ran gpt-5.5
through the OpenAI Batch API: 7,605 requests completed with 0 parse/API
errors at an estimated batch-discounted cost of \$14.13 under a \$20
budget cap. I also ran Claude Sonnet 4.6 through provider Batch API mode
for the same subset; 7,604 of 7,605 outputs parsed cleanly after
retries. Local models avoid API charges, but the relevant cost becomes
runtime, hardware availability, and energy use.

\section{Implications for applied
use}\label{implications-for-applied-use}

The results suggest a simple validation step rather than a single best
model. Start with 100 to 300 hand-labeled cases from the target task.
Test one cheap API model, one flagship API model, and at least two local
models. Report the main F1 score, accuracy, MCC, and the share of
unusable responses. For imbalanced tasks, inspect rare-class recall as
well.

Researchers should test batching on the same validation sample before
production coding. In these results, 10 items per prompt usually reduces
local runtime with little average performance loss, but the failure rate
can be high for specific model-task pairs. For long, multi-class
codebooks, researchers should not batch aggressively unless retries are
built in.

All five local models fit on a 32 GB Apple Silicon MacBook, so one
laptop setup is enough hardware for the full benchmark. When paid API
use is acceptable, cheap tiers should also be tested rather than
dismissed: in this run, gpt-5.4-nano lost 0.028 mean F1 relative to
gpt-5.5 and cost about \$4 for the full 34-task benchmark.

\section{Limitations and scope}\label{limitations-and-scope}

I stress three limitations. First, the model set reflects five specific
local models that fit a 32 GB Apple Silicon setup, plus four commercial
API tiers that include cheaper and flagship options. I did not optimize
prompts separately per model. Some prompts use the source paper's
wording; others come from source codebooks, public label definitions, or
task descriptions.

Second, each model-item pair is evaluated once. The uncertainty checks
therefore capture item and task variation, not repeated-call variation,
provider-side nondeterminism, or future API drift. API endpoints can
change behavior between releases even when the model name is stable,
while local open-weight models are easier to freeze exactly.

Third, several tasks come from public datasets or replication materials.
The benchmark cannot rule out training-data exposure. I therefore read
the results as practical performance on available political-science
coding tasks, not as a contamination-free generalization test.

The benchmark studies prompt-based classification only. For large
labeled datasets with fixed label sets, fine-tuned local models,
supervised classifiers, or embedding-based methods may be cheaper and
more reliable. Local inference keeps text on the researcher's machine,
which can make sensitive or restricted text data easier to use.

\section{Conclusion}\label{conclusion}

Can researchers use local open-weight models instead of commercial APIs
for applied text classification? The answer in this benchmark is often
yes, but not by default. Local models match or exceed the best API model
on 9 of 34 tasks in the task-specific oracle comparison, and the average
best-API advantage is 0.015 F1. At the same time, API models retain
their clearest edge on complex tasks with many labels or multiple
outputs per item.

Taken together, the results point to a validation-first workflow rather
than a model ranking. Researchers should test candidate models on
labeled examples from the target task, report both performance and
unusable-output rates, and treat batching as a speed tool that requires
its own reliability check.

\section*{References}\label{references}
\addcontentsline{toc}{section}{References}

\protect\phantomsection\label{refs}

\section*{Companion materials}\label{companion-materials}
\addcontentsline{toc}{section}{Companion materials}

Prompt sources, batching details, summary CSVs, and reproduction
instructions are on the GitHub repository at
\url{https://github.com/hhilbig/polsci-open-bench}.

\begin{landscape}

\section*{Tasks}\label{tasks}
\addcontentsline{toc}{section}{Tasks}

\begingroup\fontsize{8}{10}\selectfont

\begin{longtable}[t]{l>{\raggedright\arraybackslash}p{18em}>{\raggedright\arraybackslash}p{7em}>{\raggedright\arraybackslash}p{8em}lll}
\caption{Thirty-four tasks in the benchmark. 'Type' is the report grouping used in Figure \ref{fig-family}; it is a broad annotation type rather than a source-provided task family. 'Provenance': Verbatim = prompt unchanged from source paper, Verbatim-adapted = source content with format-only changes, Derived = my prompt based on a source codebook, public label definitions, or task description. 'Prompt': short (~15-35 lines) or long codebook (~47-126 lines). This is the trait that determines local batching feasibility. Source labels correspond to full entries in the references. The CMP task uses the Halterman \& Keith (2026) domain-collapsed 7-class version of the Manifesto Project (2025) CMP/MARPOR coding scheme, not the full CMP category set. \label{tab:tasks}}\\
\toprule
Task & Description & Source & Type & Labels & Provenance & Prompt\\
\midrule
\endfirsthead
\caption[]{Thirty-four tasks in the benchmark. 'Type' is the report grouping used in Figure \ref{fig-family}; it is a broad annotation type rather than a source-provided task family. 'Provenance': Verbatim = prompt unchanged from source paper, Verbatim-adapted = source content with format-only changes, Derived = my prompt based on a source codebook, public label definitions, or task description. 'Prompt': short (~15-35 lines) or long codebook (~47-126 lines). This is the trait that determines local batching feasibility. Source labels correspond to full entries in the references. The CMP task uses the Halterman \& Keith (2026) domain-collapsed 7-class version of the Manifesto Project (2025) CMP/MARPOR coding scheme, not the full CMP category set.  }\\
\toprule
Task & Description & Source & Type & Labels & Provenance & Prompt\\
\midrule
\endhead

\endfoot
\bottomrule
\endlastfoot
Gilardi relevance & Tweets, classified as about content moderation or not. & Gilardi et al. 2023 & Relevance \& Harm & binary & Verbatim & short\\
Brandt relevance & News articles, classified as conflict- or politics-relevant or not. & Brandt et al. 2025 & Relevance \& Harm & binary & Derived & short\\
Burnham PolNLI & Premise-hypothesis pairs, classified for entailment. & Burnham et al. 2025 & Claims \& Relations & binary & Derived & short\\
Ballard incivility & Tweets by U.S. members of Congress, classified for incivility. & Ballard 2022 & Relevance \& Harm & binary & Derived & short\\
Line of Fire incivility & Political tweets, classified for incivility. & Rheault et al. 2019 & Relevance \& Harm & binary & Derived & short\\
\addlinespace
Theocharis incivility & Political tweets, classified for incivility. & Theocharis et al. 2020 & Relevance \& Harm & binary & Derived & short\\
Gilardi stance & Tweets about content moderation, classified by stance toward moderation. & Gilardi et al. 2023 & Position \& Tone & 3-class & Verbatim & short\\
SemEval stance & Tweets toward five named political targets, classified by stance. & Chae \& Davidson 2026 & Position \& Tone & 3-class & Derived & short\\
SCOTUS sentiment & Tweets about U.S. Supreme Court rulings, classified by sentiment. & Ornstein et al. 2025 & Position \& Tone & 3-class & Derived & short\\
Wesleyan ad tone & U.S. Meta political ads, classified by tone toward a candidate. & Zhang et al. 2025 & Position \& Tone & 3-class & Derived & short\\
\addlinespace
CCC protest type & News stories about U.S. protest events, classified by event type. & Halterman \& Keith 2026 & Events \& Actions & 4-class & Verbatim-adapted & long codebook\\
BFRS violence & News stories about Pakistani political violence, classified by event type. & Halterman \& Keith 2026 & Events \& Actions & 12-class & Verbatim-adapted & long codebook\\
PAPEA protest forms & Protest snippets, classified by one of seven common source protest-form labels. & Haunss et al. 2025 & Events \& Actions & 7-class & Derived & long codebook\\
ICBe event type & Crisis-event sentences, classified by sentence event type. & Douglass et al. 2024 & Events \& Actions & 5-class & Derived & short\\
CMP manifestos & Quasi-sentences from political party manifestos, classified by policy domain. & Halterman \& Keith 2026; Manifesto Project 2025 & Issues \& Topics & 7-class & Derived & long codebook\\
\addlinespace
Cross-domain topics & Political text, classified by broad policy domain. & Osnabruegge et al. 2023 & Issues \& Topics & 8-class & Derived & long codebook\\
Campaign policy area & Campaign statements, classified by policy area. & Müller \& Fujimura 2025 & Issues \& Topics & 12-class & Derived & long codebook\\
BES MII & Open-ended British Election Study responses, classified by issue topic. & Mellon et al. 2024 & Issues \& Topics & 50-class & Verbatim-adapted & long codebook\\
Spanish protest toxicity & Spanish protest tweets, classified for toxic content. & González-Bustamante 2024 & Relevance \& Harm & binary & Derived & short\\
TwitCivility impoliteness & Political tweets, classified for impoliteness. & Pendzel et al. 2023 & Relevance \& Harm & binary & Derived & short\\
\addlinespace
GTD attack type & Global Terrorism Database events, classified by primary attack type. & Brandt et al. 2025 & Events \& Actions & 9-class & Derived & long codebook\\
PAPEA protest claims & Protest snippets, classified by source protest-claim label. & Haunss et al. 2025 & Issues \& Topics & 28-class & Derived & long codebook\\
PLOVER CAMEO event & Gold-standard CAMEO examples, classified into PLOVER event categories. & Halterman et al. 2023 & Events \& Actions & 18-class & Derived & long codebook\\
PolNLI event entailment & Event-extraction PolNLI pairs, classified for entailment. & Burnham et al. 2025 & Claims \& Relations & binary & Derived & short\\
Women's March stance & Tweets about the Women's March, classified by stance. & Bestvater \& Monroe 2023 & Position \& Tone & binary & Derived & short\\
\addlinespace
Trump stance & Political statements, classified by stance toward Trump. & Burnham 2025 & Position \& Tone & 3-class & Derived & short\\
COVID threat minimization & Political text, classified for COVID threat minimization. & Burnham 2025 & Claims \& Relations & binary & Derived & short\\
Kavanaugh stance & Tweets about Brett Kavanaugh, classified by stance. & Bestvater \& Monroe 2023 & Position \& Tone & binary & Derived & short\\
ATI topics & Mexican access-to-information requests, classified by request topic. & Erlich et al. 2022 & Issues \& Topics & 7-label & Derived & long codebook\\
CAP party platforms & Party-platform quasi-statements, classified by CAP major topic. & Policy Agendas Project 2025 & Issues \& Topics & 21-class & Derived & long codebook\\
\addlinespace
CAP CRS reports & Congressional Research Service titles and summaries, classified by CAP major topic. & Policy Agendas Project 2025; CRS Products & Issues \& Topics & 21-class & Derived & long codebook\\
PolitiCAUSE causal relation & Political sentences, classified for whether they express a causal relation. & Garcia Corral et al. 2024 & Claims \& Relations & binary & Derived & short\\
Manifesto populism & Italian manifesto sentences, classified for populist rhetoric. & Di Cocco \& Monechi 2022 & Claims \& Relations & binary & Derived & short\\
AgoraSpeech criticism/agenda & Greek campaign-speech paragraphs, classified as criticism or agenda setting. & Sermpezis et al. 2026 & Claims \& Relations & 2-class & Derived & short\\*
\end{longtable}
\endgroup{}

\end{landscape}

\section*{Per-task winners}\label{per-task-winners}
\addcontentsline{toc}{section}{Per-task winners}

\begin{table}[H]
\centering
\caption{Top model per task by the main F1 score across the unified 34-task benchmark, grouped by the report's five annotation types. Accuracy and MCC refer to the same model-task pair. CIs and the full per-(task, model) table are on the GitHub repository. \label{tab:winners}}
\centering
\fontsize{8.5}{10.5}\selectfont
\begin{tabular}[t]{llllll}
\toprule
Type & Task & Top model & Main F1 & Accuracy & MCC\\
\midrule
 & Ballard incivility & Claude Sonnet 4.6 & 0.563 & 0.848 & 0.515\\
\cmidrule{2-6}
 & Brandt relevance & Qwen3 14B & 0.739 & 0.887 & 0.703\\
\cmidrule{2-6}
 & Gilardi relevance & gpt-5.5 & 0.950 & 0.954 & 0.908\\
\cmidrule{2-6}
 & Line of Fire incivility & DeepSeek V4 Pro & 0.607 & 0.886 & 0.561\\
\cmidrule{2-6}
 & Spanish protest toxicity & Gemma 4 26B & 0.894 & 0.892 & 0.784\\
\cmidrule{2-6}
 & Theocharis incivility & gpt-5.5 & 0.682 & 0.812 & 0.551\\
\cmidrule{2-6}
\multirow[t]{-7}{*}{\raggedright\arraybackslash Relevance \& Harm} & TwitCivility impoliteness & Gemma 4 31B & 0.631 & 0.794 & 0.503\\
\cmidrule{1-6}
 & Gilardi stance & Claude Sonnet 4.6 & 0.568 & 0.648 & 0.471\\
\cmidrule{2-6}
 & Kavanaugh stance & gpt-5.5 & 0.954 & 0.952 & 0.906\\
\cmidrule{2-6}
 & SCOTUS sentiment & DeepSeek V4 Pro & 0.668 & 0.718 & 0.543\\
\cmidrule{2-6}
 & SemEval 2016 stance & gpt-5.5 & 0.773 & 0.768 & 0.659\\
\cmidrule{2-6}
 & Trump stance & Claude Sonnet 4.6 & 0.804 & 0.826 & 0.730\\
\cmidrule{2-6}
 & Wesleyan ad tone & gpt-5.4-nano & 0.711 & 0.918 & 0.808\\
\cmidrule{2-6}
\multirow[t]{-7}{*}{\raggedright\arraybackslash Position \& Tone} & Women's March stance & Claude Sonnet 4.6 & 0.905 & 0.852 & 0.623\\
\cmidrule{1-6}
 & BFRS violence (12 classes) & gpt-5.5 & 0.693 & 0.730 & 0.691\\
\cmidrule{2-6}
 & CCC protest type & Qwen3 14B & 0.499 & 0.610 & 0.387\\
\cmidrule{2-6}
 & GTD attack type (9 classes) & gpt-5.4-nano & 0.617 & 0.750 & 0.656\\
\cmidrule{2-6}
 & ICBe event type (5 classes) & Gemma 4 26B & 0.530 & 0.584 & 0.463\\
\cmidrule{2-6}
 & PAPEA protest forms (7 classes) & gpt-5.5 & 0.965 & 0.972 & 0.958\\
\cmidrule{2-6}
\multirow[t]{-6}{*}{\raggedright\arraybackslash Events \& Actions} & PLOVER CAMEO event (18 classes) & gpt-5.5 & 0.654 & 0.699 & 0.682\\
\cmidrule{1-6}
 & AgoraSpeech criticism/agenda & Mistral Sm 24B & 0.922 & 0.924 & 0.843\\
\cmidrule{2-6}
 & Burnham PolNLI & gpt-5.5 & 0.908 & 0.928 & 0.853\\
\cmidrule{2-6}
 & COVID threat minimization & Mistral Sm 24B & 0.602 & 0.819 & 0.499\\
\cmidrule{2-6}
 & Manifesto populism & Mistral Sm 24B & 0.455 & 0.904 & 0.403\\
\cmidrule{2-6}
 & PolNLI event entailment & Gemma 4 31B & 0.974 & 0.972 & 0.945\\
\cmidrule{2-6}
\multirow[t]{-6}{*}{\raggedright\arraybackslash Claims \& Relations} & PolitiCAUSE causal relation & Claude Sonnet 4.6 & 0.716 & 0.772 & 0.578\\
\cmidrule{1-6}
 & ATI topics (7 labels) & Claude Sonnet 4.6 & 0.502 & NA & NA\\
\cmidrule{2-6}
 & BES MII (50 classes) & gpt-5.5 & 0.547 & 0.946 & 0.934\\
\cmidrule{2-6}
 & CAP CRS reports (21 classes) & gpt-5.5 & 0.732 & 0.774 & 0.757\\
\cmidrule{2-6}
 & CAP party platforms (21 classes) & Claude Sonnet 4.6 & 0.731 & 0.736 & 0.719\\
\cmidrule{2-6}
 & CMP manifestos (7 classes) & gpt-5.5 & 0.597 & 0.646 & 0.545\\
\cmidrule{2-6}
 & Campaign policy area (12 classes) & gpt-5.5 & 0.636 & 0.750 & 0.691\\
\cmidrule{2-6}
 & Cross-domain topics (8 classes) & Claude Sonnet 4.6 & 0.476 & 0.494 & 0.433\\
\cmidrule{2-6}
\multirow[t]{-8}{*}{\raggedright\arraybackslash Issues \& Topics} & PAPEA protest claims (28 classes) & gpt-5.5 & 0.462 & 0.636 & 0.606\\
\bottomrule
\end{tabular}
\end{table}

\section*{Task-level uncertainty for model
averages}\label{task-level-uncertainty-for-model-averages}
\addcontentsline{toc}{section}{Task-level uncertainty for model
averages}

\begin{table}[H]
\centering
\caption{Paired task-level bootstrap intervals for cross-task mean-F1 differences among the four strongest models. Each bootstrap resamples the 34 tasks with replacement and recomputes the paired mean difference. These intervals assess sensitivity to benchmark composition; they do not replace the paired-by-item bootstrap intervals used for within-task comparisons. \label{tab:task-bootstrap}}
\centering
\fontsize{9}{11}\selectfont
\begin{tabular}[t]{lrrr}
\toprule
Model comparison & Mean difference & 95\% task-bootstrap interval & Tasks\\
\midrule
Claude Sonnet 4.6 - gpt-5.5 & 0.008 & {}[-0.008, 0.024] & 34\\
Claude Sonnet 4.6 - Gemma 4 31B & 0.020 & {}[0.005, 0.038] & 34\\
Claude Sonnet 4.6 - DeepSeek V4 Pro & 0.021 & {}[0.004, 0.040] & 34\\
gpt-5.5 - Gemma 4 31B & 0.012 & {}[-0.007, 0.033] & 34\\
gpt-5.5 - DeepSeek V4 Pro & 0.013 & {}[-0.006, 0.032] & 34\\
\addlinespace
Gemma 4 31B - DeepSeek V4 Pro & 0.001 & {}[-0.020, 0.020] & 34\\
\bottomrule
\end{tabular}
\end{table}

\section*{Local runtime per 1,000
items}\label{local-runtime-per-1000-items}
\addcontentsline{toc}{section}{Local runtime per 1,000 items}

\begin{figure}[!h]

\centering{

\includegraphics[width=0.82\linewidth,height=\textheight,keepaspectratio]{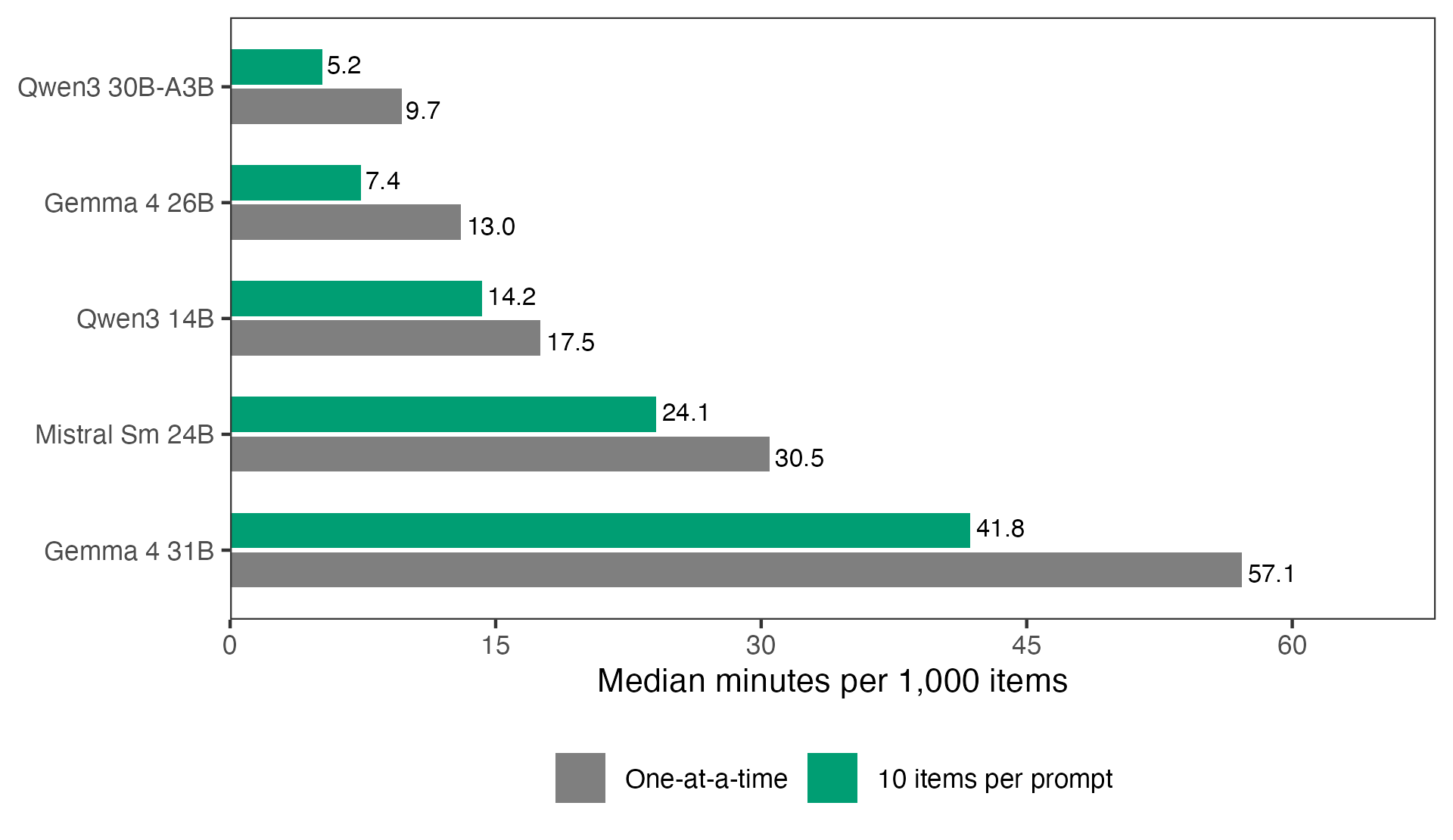}

}

\caption{\label{fig-local-runtime-per-1000}Median local runtime per
1,000 items. One-at-a-time bars and bars for 10 items per prompt use the
34-task benchmark. Lower values are faster. Batching should be
interpreted alongside response-format and label failure checks.}

\end{figure}

Figure \ref{fig-local-runtime-per-1000} converts local median runtime
per item into minutes per 1,000 items. The bars for 10 items per prompt
use the full 34-task local batching grid, so they should be read
alongside the response-format and label failure checks below.

\section*{Model-selection upper
bounds}\label{model-selection-upper-bounds}
\addcontentsline{toc}{section}{Model-selection upper bounds}

Figure \ref{fig-best-local-api-gap} in the main text compares the best
local model with the best API model separately for each task. The
largest API advantages appear on tasks that involve policy domains,
event codes, causal relations, or protest claims. The largest local
advantages appear on more focused tasks such as manifesto populism,
COVID threat minimization, and CCC protest type.

\begin{table}[H]
\centering
\caption{Mean F1 under single-model choices and task-specific upper bounds. The task-specific rows choose the best model after observing task-level performance, so they describe an upper bound rather than expected performance from a validation sample. \label{tab:model-selection-value}}
\centering
\fontsize{9}{11}\selectfont
\begin{tabular}[t]{llr}
\toprule
Selection rule & Model or rule & Mean F1\\
\midrule
Best single model & Claude Sonnet 4.6 & 0.668\\
Best single local model & Gemma 4 31B & 0.648\\
Best single API model & Claude Sonnet 4.6 & 0.668\\
Task-specific upper bound: local & Best local per task & 0.673\\
Task-specific upper bound: API & Best API per task & 0.688\\
\addlinespace
Task-specific upper bound: all models & Best model per task & 0.696\\
\bottomrule
\end{tabular}
\end{table}

The best single model is Claude Sonnet 4.6, with a 34-task mean F1 of
0.668. The best single local model, Gemma 4 31B, reaches 0.648. If the
local model can vary by task, the local task-specific upper bound rises
to 0.673. The API upper bound reaches 0.688, and the all-model upper
bound reaches 0.696. Because these rows choose after observing outcomes,
they describe an upper bound on what validation can buy, not expected
production performance.

\section*{Response-format or label failure rates above 5
percent}\label{response-format-or-label-failure-rates-above-5-percent}
\addcontentsline{toc}{section}{Response-format or label failure rates
above 5 percent}

\begin{table}[H]
\centering
\caption{Batched model-task pairs with response-format or label failure rates of at least 5 percent, not a complete list of all model-task pairs. Failures include responses that do not parse, responses with the wrong object or array shape, and responses that use labels outside the allowed label set. Local rows come from the full 34-task grid with 10 items per prompt for the five local models (170 pairs). OpenAI rows come from a separate 10-task prompt-batching diagnostic with 10 and 20 items per prompt (40 pairs); these rows do not enter the main one-item-at-a-time benchmark. Claude Sonnet and DeepSeek API prompt batching are not covered here. \label{tab:malformed}}
\centering
\fontsize{8}{10}\selectfont
\begin{tabular}[t]{llrlr}
\toprule
Task & Model & Batch size & Scope & Failure rate\\
\midrule
ATI topics (7 labels) & Gemma 4 26B & 10 & Local b=10 & 72\%\\
CCC protest & gpt-5.5 & 10 & OpenAI diagnostic & 68\%\\
CCC protest type & Qwen3 30B-A3B & 10 & Local b=10 & 51\%\\
Wesleyan ads & gpt-5.5 & 20 & OpenAI diagnostic & 48\%\\
CCC protest & gpt-5.5 & 20 & OpenAI diagnostic & 40\%\\
\addlinespace
BFRS Pakistan & gpt-5.5 & 20 & OpenAI diagnostic & 32\%\\
Cross-domain topics (8 classes) & Qwen3 30B-A3B & 10 & Local b=10 & 30\%\\
BFRS Pakistan & gpt-5.5 & 10 & OpenAI diagnostic & 28\%\\
Wesleyan ads & gpt-5.5 & 10 & OpenAI diagnostic & 26\%\\
BFRS violence (12 classes) & Qwen3 30B-A3B & 10 & Local b=10 & 18\%\\
\addlinespace
Ballard incivility & gpt-5.5 & 20 & OpenAI diagnostic & 16\%\\
SemEval stance & gpt-5.5 & 20 & OpenAI diagnostic & 16\%\\
Wesleyan ad tone & Qwen3 30B-A3B & 10 & Local b=10 & 16\%\\
BFRS Pakistan & gpt-5.4-nano & 10 & OpenAI diagnostic & 12\%\\
SCOTUS sentiment & gpt-5.5 & 20 & OpenAI diagnostic & 12\%\\
\addlinespace
Ballard incivility & gpt-5.5 & 10 & OpenAI diagnostic & 10\%\\
CMP manifestos & gpt-5.5 & 10 & OpenAI diagnostic & 10\%\\
GTD attack type (9 classes) & Qwen3 30B-A3B & 10 & Local b=10 & 10\%\\
CCC protest & gpt-5.4-nano & 20 & OpenAI diagnostic & 8\%\\
BFRS Pakistan & gpt-5.4-nano & 20 & OpenAI diagnostic & 8\%\\
\addlinespace
PLOVER CAMEO event (18 classes) & Qwen3 30B-A3B & 10 & Local b=10 & 6\%\\
CCC protest & gpt-5.4-nano & 10 & OpenAI diagnostic & 6\%\\
\bottomrule
\end{tabular}
\end{table}

\section*{Performance under alternative
metrics}\label{performance-under-alternative-metrics}
\addcontentsline{toc}{section}{Performance under alternative metrics}

\begin{table}[H]
\centering
\caption{Unified 34-task model averages. F1 coverage is complete for all nine models. Accuracy and MCC are averaged over tasks where the metric is defined in the same way across outputs; the multi-binary ATI task is excluded from those two columns. The main F1 score is the task-specific summary metric. \label{tab:altmetrics}}
\centering
\fontsize{9}{11}\selectfont
\begin{tabular}[t]{lrrrrrr}
\toprule
Model & F1 tasks & Accuracy tasks & MCC tasks & Mean F1 & Mean accuracy & Mean MCC\\
\midrule
Claude Sonnet 4.6 & 34/34 & 33/34 & 33/34 & 0.668 & 0.777 & 0.632\\
gpt-5.5 & 34/34 & 33/34 & 33/34 & 0.660 & 0.775 & 0.627\\
Gemma 4 31B & 34/34 & 33/34 & 33/34 & 0.648 & 0.770 & 0.615\\
DeepSeek V4 Pro & 34/34 & 33/34 & 33/34 & 0.647 & 0.772 & 0.613\\
Gemma 4 26B & 34/34 & 33/34 & 33/34 & 0.639 & 0.762 & 0.597\\
\addlinespace
gpt-5.4-nano & 34/34 & 33/34 & 33/34 & 0.632 & 0.748 & 0.591\\
Mistral Sm 24B & 34/34 & 33/34 & 33/34 & 0.629 & 0.740 & 0.583\\
Qwen3 14B & 34/34 & 33/34 & 33/34 & 0.608 & 0.749 & 0.570\\
Qwen3 30B-A3B & 34/34 & 33/34 & 33/34 & 0.593 & 0.744 & 0.555\\
\bottomrule
\end{tabular}
\end{table}

\end{document}